\documentclass[aps,prl,twocolumn,superscriptaddress,showpacs]{revtex4-1}
\usepackage{amsmath,amssymb,amsfonts,amsthm}
\usepackage{graphicx}
\usepackage{bbm}
\usepackage{dcolumn}   
\usepackage{epstopdf}
\usepackage[colorlinks=true,linkcolor=blue,urlcolor=blue,citecolor=blue]{hyperref}
\usepackage[caption=false]{subfig}

\begin{document}

 \newcommand{\breite}{1.0} 

\newtheorem{prop}{Proposition}
\newtheorem{cor}{Corollary}

\newcommand{\be}{\begin{equation}}
\newcommand{\ee}{\end{equation}}

\newcommand{\bea}{\begin{eqnarray}}
\newcommand{\eea}{\end{eqnarray}}

\newcommand{\Reals}{\mathbb{R}}     
\newcommand{\Com}{\mathbb{C}}       
\newcommand{\Nat}{\mathbb{N}}       

\newcommand{\id}{\mathbbm{1}}    

\newcommand{\Real}{\mathop{\mathrm{Re}}}
\newcommand{\Imag}{\mathop{\mathrm{Im}}}

\def\O{\mbox{$\mathcal{O}$}}   
\def\F{\mathcal{F}}			
\def\sgn{\text{sgn}}

\newcommand{\deo}{\ensuremath{\Delta_0}}
\newcommand{\dea}{\ensuremath{\Delta}}
\newcommand{\ak}{\ensuremath{a_k}}
\newcommand{\ad}{\ensuremath{a^{\dagger}_{-k}}}
\newcommand{\sx}{\ensuremath{\sigma_x}}
\newcommand{\sz}{\ensuremath{\sigma_z}}
\newcommand{\spl}{\ensuremath{\sigma_{+}}}
\newcommand{\smi}{\ensuremath{\sigma_{-}}}
\newcommand{\alk}{\ensuremath{\alpha_{k}}}
\newcommand{\bk}{\ensuremath{\beta_{k}}}
\newcommand{\ok}{\ensuremath{\omega_{k}}}
\newcommand{\vd}{\ensuremath{V^{\dagger}_1}}
\newcommand{\vi}{\ensuremath{V_1}}
\newcommand{\vo}{\ensuremath{V_o}}
\newcommand{\zc}{\ensuremath{\frac{E_z}{E}}}
\newcommand{\xc}{\ensuremath{\frac{\Delta}{E}}}
\newcommand{\xd}{\ensuremath{X^{\dagger}}}
\newcommand{\aok}{\ensuremath{\frac{\alk}{\ok}}}
\newcommand{\tpw}{\ensuremath{e^{i \ok s }}}
\newcommand{\tpe}{\ensuremath{e^{2iE s }}}
\newcommand{\tmw}{\ensuremath{e^{-i \ok s }}}
\newcommand{\tme}{\ensuremath{e^{-2iE s }}}
\newcommand{\epls}{\ensuremath{e^{F(s)}}}
\newcommand{\emis}{\ensuremath{e^{-F(s)}}}
\newcommand{\epl}{\ensuremath{e^{F(0)}}}
\newcommand{\emi}{\ensuremath{e^{F(0)}}}

\newcommand{\mkcomm}[1]{{\color{red}MK: #1}}

\newcommand{\lr}[1]{\left( #1 \right)}
\newcommand{\lrs}[1]{\left( #1 \right)^2}
\newcommand{\lrb}[1]{\left< #1\right>}
\newcommand{\nbt}{\ensuremath{\lr{ \lr{n_k + 1} \tmw + n_k \tpw  }}}

\newcommand{\om}{\ensuremath{\omega}}
\newcommand{\dw}{\ensuremath{\Delta_0}}
\newcommand{\wbp}{\ensuremath{\omega_0}}
\newcommand{\dv}{\ensuremath{\Delta_0}}
\newcommand{\vbp}{\ensuremath{\nu_0}}
\newcommand{\vplus}{\ensuremath{\nu_{+}}}
\newcommand{\vminus}{\ensuremath{\nu_{-}}}
\newcommand{\wplus}{\ensuremath{\omega_{+}}}
\newcommand{\wminus}{\ensuremath{\omega_{-}}}
\newcommand{\uv}[1]{\ensuremath{\mathbf{\hat{#1}}}} 
\newcommand{\abs}[1]{\left| #1 \right|} 
\newcommand{\avg}[1]{\left< #1 \right>} 
\let\underdot=\d 
\renewcommand{\d}[2]{\frac{d #1}{d #2}} 
\newcommand{\dd}[2]{\frac{d^2 #1}{d #2^2}} 
\newcommand{\pd}[2]{\frac{\partial #1}{\partial #2}} 
\newcommand{\pdd}[2]{\frac{\partial^2 #1}{\partial #2^2}} 
\newcommand{\pdc}[3]{\left( \frac{\partial #1}{\partial #2}
 \right)_{#3}} 
\newcommand{\ket}[1]{\left| #1 \right>} 
\newcommand{\bra}[1]{\left< #1 \right|} 
\newcommand{\braket}[2]{\left< #1 \vphantom{#2} \right|
 \left. #2 \vphantom{#1} \right>} 
\newcommand{\matrixel}[3]{\left< #1 \vphantom{#2#3} \right|
 #2 \left| #3 \vphantom{#1#2} \right>} 
\newcommand{\grad}[1]{\gv{\nabla} #1} 
\let\divsymb=\div 
\renewcommand{\div}[1]{\gv{\nabla} \cdot #1} 
\newcommand{\curl}[1]{\gv{\nabla} \times #1} 
\let\baraccent=\= 

\title{Anomalous diffusion and Griffiths effects near the many-body localization transition}

\author{Kartiek Agarwal}
\affiliation{Physics Department, Harvard University, Cambridge, Massachusetts 02138, USA}
\email[]{agarwal@physics.harvard.edu}
\author{Sarang Gopalakrishnan}
\affiliation{Physics Department, Harvard University, Cambridge, Massachusetts 02138, USA}
\author{Michael Knap}
\affiliation{Physics Department, Harvard University, Cambridge, Massachusetts 02138, USA}
\affiliation{ITAMP, Harvard-Smithsonian Center for Astrophysics, Cambridge, MA 02138, USA}
\author{Markus M\"uller}
\affiliation{The Abdus Salam International Center for Theoretical Physics, Strada Costiera 11, 34151 Trieste, Italy}
\author{Eugene Demler}
\affiliation{Physics Department, Harvard University, Cambridge, Massachusetts 02138, USA}

\date{\today}
\begin{abstract}

We explore the high-temperature dynamics of the disordered, one-dimensional XXZ model near the many-body localization (MBL) transition, focusing on the delocalized (i.e., ``metallic'') phase. In the vicinity of the transition, we find that this phase has the following properties: (i) Local magnetization fluctuations relax subdiffusively; (ii) the a.c. conductivity vanishes near zero frequency as a power law; (iii) the distribution of resistivities becomes increasingly broad at low frequencies, approaching a power law in the zero-frequency limit. We argue that these effects can be understood in a unified way if the metallic phase near the MBL transition is a quantum Griffiths phase. We establish scaling relations between the associated exponents, assuming a scaling form of the spin-diffusion propagator. A phenomenological classical resistor-capacitor model captures all the essential features.

\end{abstract}
\maketitle

Noninteracting electrons in disordered media display a uniquely quantum phenomenon known as Anderson localization~\cite{anderson_absence_1958}; when all electronic states are Anderson localized, dc transport is absent. Evidence from perturbative~\cite{fleishman,basko_metalinsulator_2006, gornyi_interacting_2005}, numerical~\cite{oganesyan_localization_2007, monthus_many-body_2010, berkelbach_conductivity_2010, pal_many-body_2010}, and rigorous mathematical approaches~\cite{imbrie} suggests that the main features of Anderson localization (in particular, the absence of diffusion and dc transport) persist in the presence of interactions. The resulting phase, known as the many-body localized (MBL) phase~\cite{arcmp}, has a number of remarkable features: a system in the MBL phase is non-ergodic---i.e., its many-body eigenstates violate the eigenstate 
thermalization hypothesis~\cite{pal_many-body_2010,luca_ergodicity_2013,laumann_manybody_2014}---and supports extensively many local conserved quantities~\cite{vosk_many-body_2013,huse_phenomenology_2013,serbyn_local_2013,swingle_simple_2013,ros_integrals_2014,chandran_constructing_2014}. 
Consequences of MBL  such as slow entanglement growth~\cite{znidaric_many-body_2008,bardarson_unbounded_2012,vosk_many-body_2013,serbyn_universal_2013} and unconventional phase transitions~\cite{huse_localization_2013,pekker_hilbert-glass_2014,chandran_many-body_2014,bauer_area_2013,bahri_localization_2013,vosk_dynamical_2014,kjall_many-body_2014,nandkishore_marginal_2014} have been analyzed and their experimental implications discussed~\cite{yao_manybody_2013,nandkishore_spectral_2014, serbyn_interfero_2014,gopalakrishnan_mean-field_2014,johri_numerical_2014,vasseur_quantum_2014}.

\begin{figure}[b]
\includegraphics[width=3.2in]{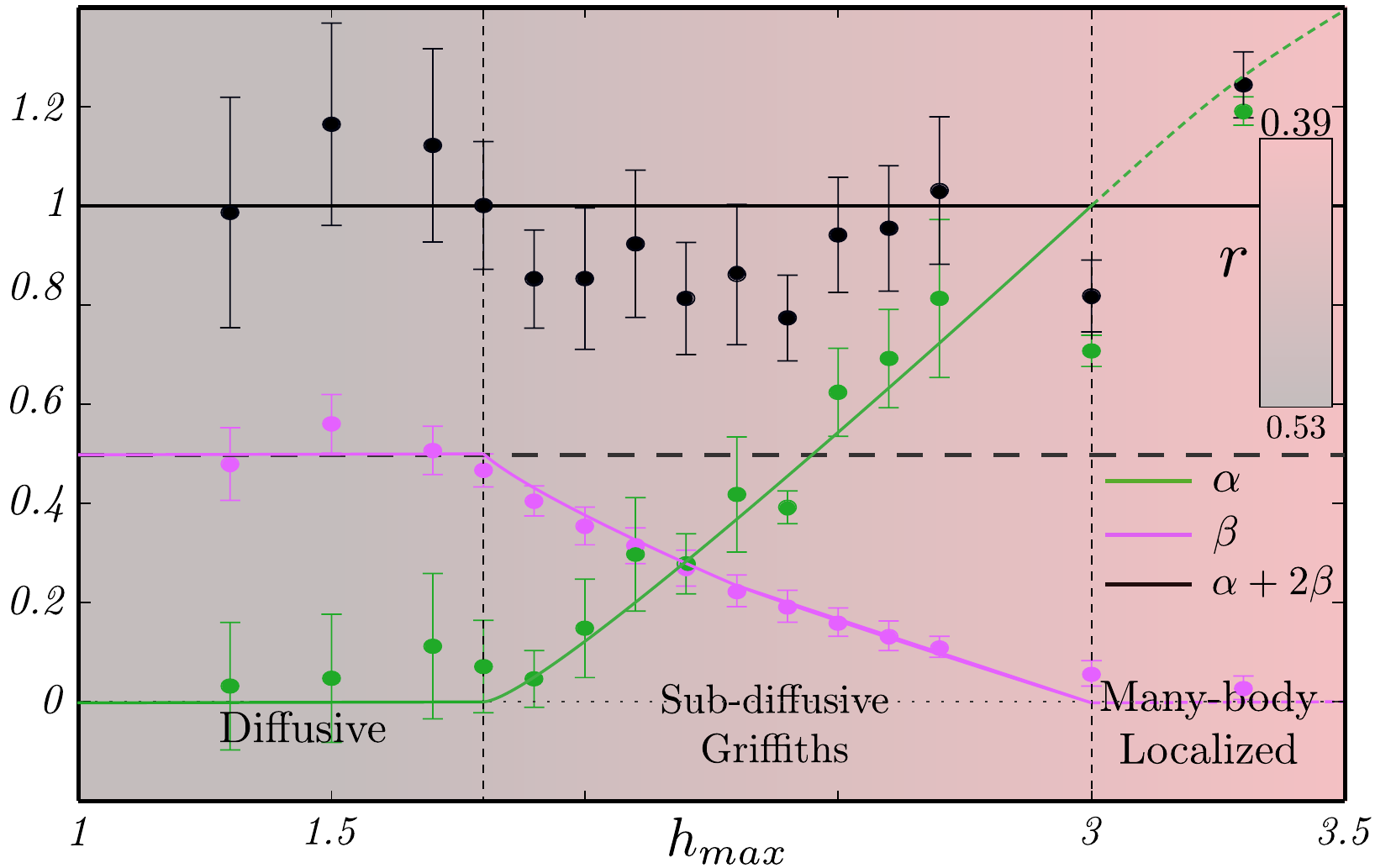}
\caption{The phase diagram of the random field, XXZ model for $J_z = 1$. $h_\text{max}$ characterizes disorder strength. As the disorder is increased, the system transitions smoothly into a sub-diffusive Griffiths-like phase with an anomalous diffusion exponent $\beta$ and exponent $\alpha$ characterizing low-frequency optical conductivity, which satisfy the scaling relation $\alpha + 2 \beta = 1$. The MBL transition is predicted to occur where $\sigma(\omega) \sim \omega$. Within precision, it coincide with the transition point determined from the level statistics parameter $r$ (see main text).}
\label{fig:phase}
\end{figure} 

While there has been a great deal of recent work establishing the existence and properties of the MBL phase, little is known about the transition between the MBL and delocalized phases. It is expected that, for sufficiently weak disorder and strong interactions~\cite{basko_metalinsulator_2006, gornyi_interacting_2005}, eigenstates should remain ergodic and transport should be diffusive, as in clean nonintegrable systems~\cite{mukerjee2006}. However, it has been proposed that diffusivity and/or ergodicity may break down as the MBL transition is approached~\cite{randomgraph, grover}, even before transport vanishes: thus, there might be an intermediate phase, or phases, between the conventional metallic phase and the MBL phase. 

In this Letter, we provide numerical evidence that an intermediate, non-diffusive phase, indeed exists. To this end, we examine the dynamical properties of 
the random-field, spin-1/2 XXZ chain at intermediate disorder strengths (i.e., in the vicinity of the MBL transition), using exact diagonalization. 
In particular, we examine the infinite-temperature, low-frequency behavior of the optical conductivity $\sigma(\omega)$, and the long-time dependence of the return probability
$C_{zz} (t)$. These probes are complementary: $\sigma(\omega)$ probes long-wavelength behavior, while $C_{zz} (t)$ is a local probe. 

Our numerical results indicate that both of these quantities exhibit anomalous power laws that vary smoothly as a function of the disorder strength in this intermediate regime. Specifically, $\sigma(\omega) \sim \omega^\alpha$ and $C_{zz}(t) \sim t^{-\beta}$, with the scaling relation $\alpha + 2\beta = 1$ [Fig.~\ref{fig:phase}]. 
Furthermore, we compute the full distribution  $D[\rho(\omega)]$ of resistivities $\rho$ at a fixed sample size as a function of frequency; we find that the width of this distribution \emph{diverges} in the low-frequency limit as $\Delta \rho (\omega) \sim 1/\omega^{\alpha'}$. Such behavior is characteristic of a quantum Griffiths phase~\cite{Vojta_griffiths}, in which power-law correlations emerge due to the interplay between the exponential rareness of large insulating regions and their exponentially large resistance. 
We account for these scaling relations by postulating a scaling form of the spin-diffusion propagator and a phenomenological resistor-capacitor model with power-law-distributed resistors.

As our work was nearing completion, a related numerical study by Bar-Lev et al.~\cite{reichman2014} appeared. While our numerical results are consistent with those of Ref.~\cite{reichman2014}, we are also able to provide an analytic understanding of the subdiffusive phase (see also Ref.~\cite{vosk2014theory}). 
\begin{figure*}
\includegraphics[width=.98\textwidth]{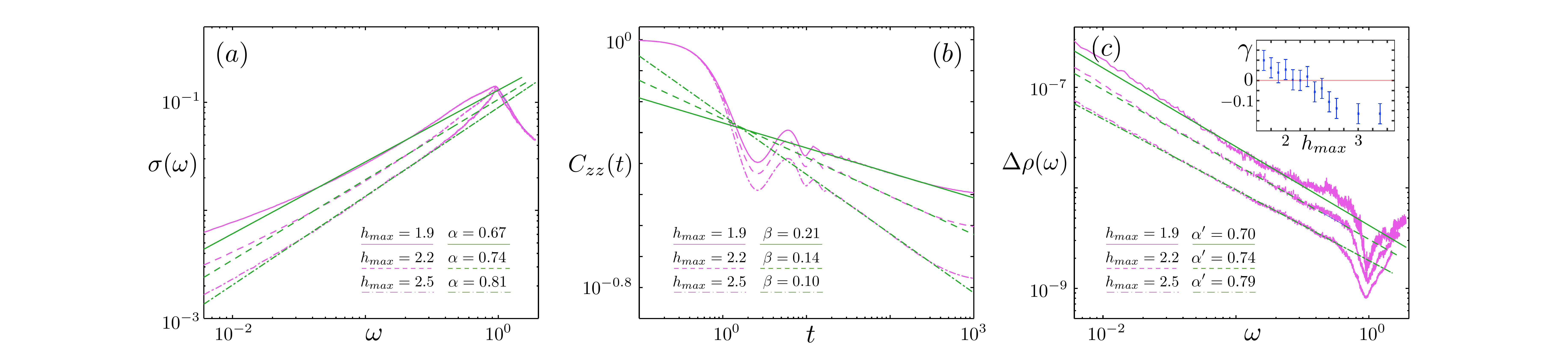}
\caption{ Behavior of (a) optical conductivity $\sigma(\omega)$, (b) return probability $C_{zz} (t)$, and (c) width $\Delta \rho(\omega)$ of the distribution of resistivities as a function of frequency (magenta). All plots are for $J_z = 0.8$, and disorder strengths as indicated in the legend. The fits (green) are power laws of the form $C_{zz} \sim 1/t^\beta$, $\sigma(\omega) \sim \omega^\alpha$, and $\Delta \rho(\omega) \sim 1/\omega^{\alpha'}$. The inset in (c) shows the relative power $\gamma = \alpha' - \alpha$ governing the scaling of the ratio $\Delta \rho(\omega) / \avg{\rho(\omega)}$ of the width and the mean of the resistivity distribution.}
\label{fig:scaling}
\end{figure*}

\paragraph*{Numerical Simulations.} We work with the XXZ model given by the Hamiltonian
\be
H = \sum_i h_i S^z_i +  \sum_{\langle ij \rangle} J (S^x_i S^x_j +  S^y_i S^y_j) + J_z S^z_i S^z_j , 
\ee
where $\langle i j\rangle$ implies sites $i$ and $j$ are nearest neighbors. The local magnetic field values $h_i$ are picked uniformly from the range $[-h_\text{max}, h_\text{max}]$; $h_\text{max}$ characterizes the strength of the disorder. The exponents $\alpha, \beta$ presented in Fig.\ref{fig:phase} were extrapolated from finite size results computed using system sizes $L = 12,14,16$ (see Supplemental Material), while results in Fig. \ref{fig:scaling} correspond to $L = 14$. We use $J=1$ as unit of energy, and choose the interaction strength to be close to the Heisenberg point, $J_z \lesssim 1$, as finite-size effects are more severe for $J_z/J \ll 1$.
The XXZ chain is expected to exhibit an infinite-temperature transition to the MBL phase at a critical $h_\text{max}$~\cite{oganesyan_localization_2007}. In what follows, we restrict ourselves to infinite temperature and choose the subspace of total magnetization $\sum_i S^z_i = 0$. 

The real part of the optical conductivity $\sigma (\omega)$ in linear response reads 
\begin{align}
T \sigma (\omega) &= \frac{T}{L} \frac{1 - e^{-\omega/T}}{\omega} \sum_{m n} \frac{e^{-\beta E_n}}{Z} \; | \langle m | \sum_i j_i | n \rangle|^2 \; \delta(\omega - \omega_{mn}) \nonumber \\
& \stackrel{T \rightarrow \infty}{\approx} \frac{1}{LZ} \sum_{m n} | \langle m | \sum_i  j_i | n \rangle|^2 \delta(\omega - \omega_{mn})
\label{eq:cond}
\end{align}
where $m,n$ are the many-body eigenstates of the system with energies $E_m,E_n$, which we evaluate using exact diagonalization, $\omega_{mn} = E_m - E_n$, and $T$ is the temperature (we set $\hbar = k_B = 1$). The first line of Eq.~(\ref{eq:cond}) is the Lehmann representation of $T\sigma (\omega)$, given in terms of a sum over local current operators $j_i$, which are 
related to the spin operators using the continuity equations, $j_i - j_{i + 1} = \partial_t S^z_i$. The second line  of Eq.~(\ref{eq:cond}) is the limiting behavior of $T\sigma(\omega)$ as $T \rightarrow \infty$. In the remainder of the manuscript the factor $T$ is implicitly understood when we refer to the conductivity $\sigma(\omega)$.
In our numerics, we use a Lorentzian form for the $\delta$-function with a width $\eta \sim \Delta/10^2$, where $\Delta = h_\text{max} \sqrt{L} / 2^L$ is approximately the average level spacing $ \sim 10^{-3} - 10^{-2}$ for the system size $L \sim 14$ and disorder strengths $h_\text{max} \sim 1.5 - 3.5$ that we explore. The precise value of $\eta$ is unimportant, so long as it is appreciably smaller than $\Delta$ (see Supplemental Material).The return probability, $C^i_{zz} (t)$, is defined as $C^i_{zz} (t) = 4 \langle S^z_i (t) S^z_i (0) \rangle$, where $i$ is any site on the chain. 
 Since we are interested in describing the phase close to the MBL transition, we also require an additional, independent, method to identify the transition point.  
 
Following Ref.~\cite{oganesyan_localization_2007} we consider the level statistics parameter $r_m= \delta^m_{-} / \delta^m_{+}$, 
where $\delta^m_\pm$ are the energy differences between eigenstate $m$ and the two adjacent eigenstates with $\delta^m_-<\delta^m_+$. The average over all eigenstates $m$, $r = \avg{r_m}$, is known to assume different values, $r \sim 0.39$ and $r \sim 0.53$ in the cases of the MBL and the conducting phase, respectively. We crudely estimate the MBL transition as the point when $r$ is halfway between these values, as determined for a $L = 16$ system (dashed line at $h_\text{max} \approx 3$ in Fig.~\ref{fig:phase}).

\paragraph*{Numerical results.} Our numerical results on the dynamic observables are summarized in Fig.~\ref{fig:scaling}. Both the optical conductivity and the return probability obey power-law behavior over multiple decades. For the exponenents defined by $\sigma(\omega)\sim \omega^\alpha$ and $C_{zz} (t)\sim 1/t^\beta$, respectively, we numerically find the scaling relation $\alpha + 2 \beta \approx 1$; the physical origin of this relation is discussed below. 
This scaling relation also holds in the diffusive regime, where one expects that $C_{zz} (t) \sim 1/\sqrt{t}$ and $\sigma(\omega) \sim \textrm{constant}$ [thus $\alpha = 0, \beta = 1/2$]. Our results approach diffusive values at small disorder $h_{max} \approx 1.5$, but we are unable to extract reliable power laws in this limit. 
As the system approaches the MBL transition, $\alpha$ continuously increases toward 1 and $\beta$ decreases to 0; $\beta = 0$ implies the absence of relaxation, and marks the transition into the MBL phase. The trend suggests that at the MBL transition $\sigma(\omega) \sim \omega$; this differs sharply from the expectation for a noninteracting Anderson insulator~\cite{mott1968conduction}, viz. $\sigma(\omega) \sim \omega^2 \log^2(\omega)$. 

Next, we look at the distribution $D[\rho(\omega)]$ of resistivities $\rho$, at a fixed frequency $\omega$. We find that the distribution of resistivities at a fixed sample size ($L = 14$) grows increasingly broad at low frequencies. The distribution has an exponential tail (see Supplemental Material); moreover, the standard deviation of sample resistivities at a fixed frequency $\omega$ diverges as $\omega^{-\alpha'}$, see Fig.~\ref{fig:scaling}.
At low frequencies, sufficiently high moments of the resistivity distribution become ill-defined and we expect that the distribution approaches a power law $P(\rho) \sim \rho^{-\tau}$ in this limit. While we are unable to reliably extract the exponent $\tau$ directly from the data, owing to the difficulty of taking the dc limit in a finite system, we will argue below that our numerical results can be explained using a resistor-capacitor network which predicts that $\tau = 2/(\alpha+1)$ in the Griffiths regime. 

\emph{Origin of scaling relation $\alpha + 2 \beta = 1$}. 
The origin of the relation $\alpha + 2\beta = 1$ can be understood very simply as follows. The relation between length and time in the subdiffusive phase can be written in the ``diffusive'' form $x^2 \sim D(t) t$, where $D(t)$ is a time-dependent diffusion constant. At long times, the scaling of return probability implies that $x^2 \sim t^{2\beta}$ in the subdiffusive phase. Thus, $D(t) \sim t^{2 \beta - 1}$, and from the Einstein relation $\sigma(\omega) \sim D(t = 2\pi/\omega) \sim \omega^{1 - 2\beta}$.

Much more generally, the relation $\alpha + 2 \beta = 1$ follows if one assumes that the average spin density propagator takes the scaling form $G(x, t) \sim t^{-\beta} \phi(x/t^\beta)$ [or, equivalently, $G(k,\omega) \sim \omega^{-1} \tilde{\phi}(k/\omega^\beta)$], i.e., if one stipulates that lengths and times are related exclusively through the dynamical exponent $z = 1/\beta$. A dynamical exponent that smoothly varies with disorder strength is reminiscent of the zero-temperature Griffiths phase in random magnets (see Ref.~\cite{motrunich2001dynamics} and references therein). We assume further that the static compressibility of the system evolves smoothly near the MBL transition (as expected for a high-temperature system with short-range interactions). From these assumptions it follows (Supplemental Material) that the dynamic structure factor $S(k, \omega) \sim \omega^{-1} h(k/\omega^\beta)$ (where $h$ is yet another scaling function, and $h(0)$ is finite). The dynamic structure factor is closely related 
to the momentum-dependent 
conductivity~\cite{igb}: specifically, $\sigma(k,\omega) \sim \omega^2 \partial_k^2 S(k, \omega) = \omega \partial_k^2 h(k/\omega^\beta)$. Again, it follows that $\sigma(k = 0, \omega) \sim \omega^{1 - 2 \beta}$.

\emph{Griffiths-phase interpretation}. 
The reasoning above related $\alpha$ to $\beta$, but did not account for the subdiffusive behavior itself. We now provide an interpretation of sub-diffusion in terms of Griffiths effects. Near the MBL transition, one expects the system to consist of metallic segments separated by insulating barriers, i.e., local regions where the system parameters favor localization. Barriers through which the tunneling time is $\agt t$ confine the magnetization at the timescale $t$. The scaling between length and time suggests that the average distance between such insulating barriers is $d(t) \sim t^\beta$. As long as $\beta<1/2$,  the time $t$ spent to tunnel through the insulating regions is parametrically larger than the timescale to diffuse between barriers, $t \gg d(t)^2 \sim t^{2\beta}$. The long time dynamics is therefore limited by these rare insulating regions. When approaching the diffusive limit $\beta=1/2$, the separation between barrier tunneling and diffusion time does not exist. Thus rare barriers cannot be 
defined and transport is simply governed by diffusion. From these considerations it follows, that a local charge excess decays to $1/d(t)$ in time $t$ which yields a return probability $C_{zz} (t) \sim 1/t^\beta$. 

This picture of insulating barriers also yields the correct scaling of the optical conductivity. We imagine that we apply a field $E$ that flips at a frequency $\omega \sim 1/t$: between such flips, the charge equilibrates to the (approximately) linear potential gradient set up by the field between insulating barriers separated by a length $d(t)$. Thus, at a position $x$, the charge flips between $\pm Ex/T$ in time $t$. This requires average current densities of order $\abs{j} \sim E d(t)^2 / t T$, a fraction of which is in phase with the applied field and thus gives rise to scaling of the real part of the conductivity, $\sigma(\omega) \sim \omega^{1- 2 \beta}$. 

Note that this Griffiths picture is qualitatively \emph{distinct} from the situation where only certain rare sites exhibit slow decay, while on most sites magnetization decays rapidly. In such a scenario, the sub-diffusive behavior would only show up in the average and not the \emph{typical} return probability. Instead, we find sub-diffusive decay in both typical and average correlations (Supplementary Information).  

\begin{figure}
\includegraphics[width=3.2in]{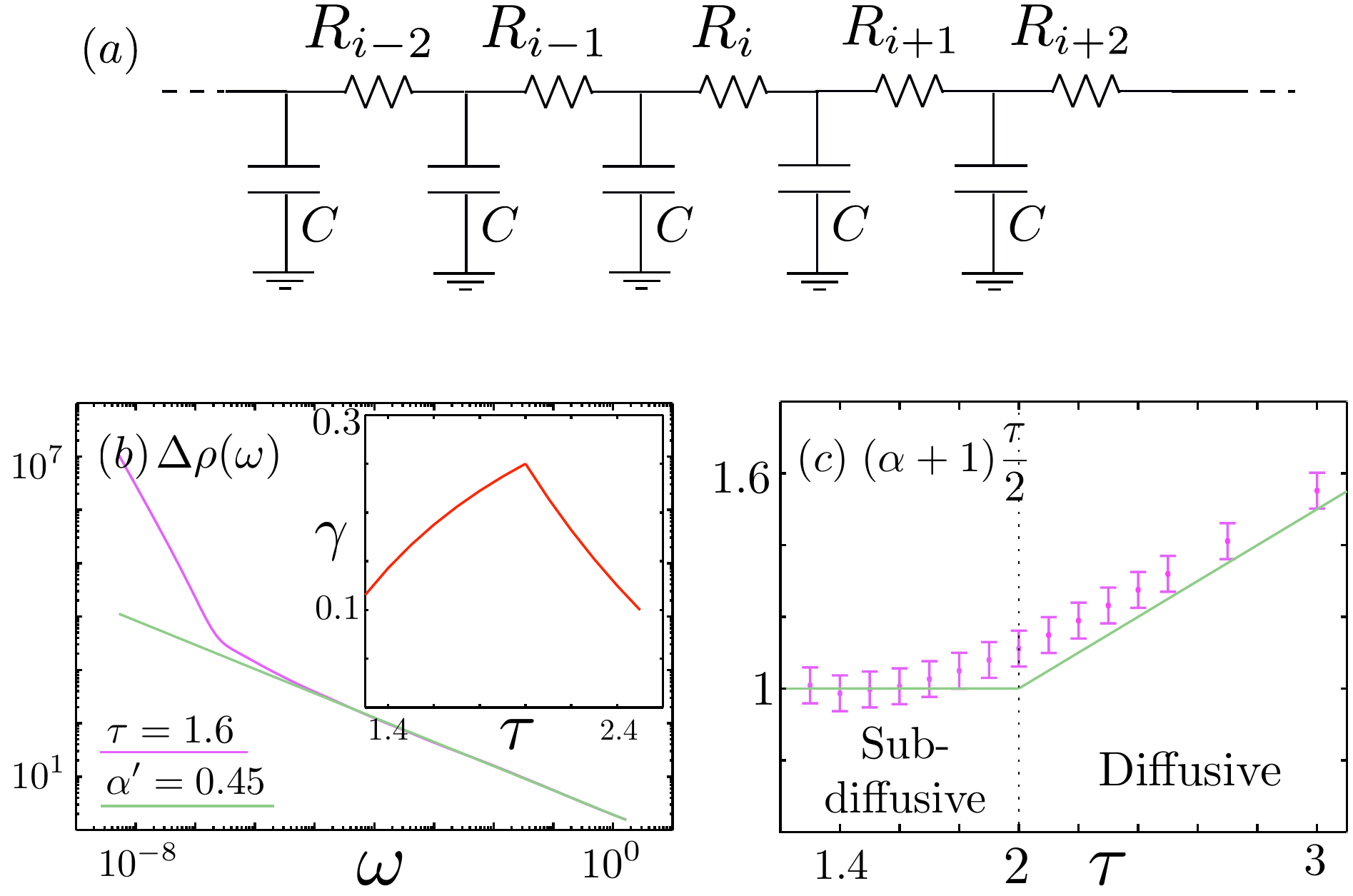}
\caption{(a) The RC model. (b) Width of the finite frequency resistivity distribution $\Delta \rho (\omega)$ (magenta) and its asymptotic form $1/\omega^{\alpha'}$ (green) plotted as a function of frequency. The inset shows $\gamma = \alpha' - \alpha \gtrsim 0$, where $\alpha$ is the exponent of the average resistivity $\bar{\rho}(\omega) \sim 1/\omega^\alpha$. (c) $(\alpha + 1) \tau/2$ as a function of $\tau$.}
\label{fig:rcscaling}
\end{figure}

\emph{RC model}. We now introduce a classical RC network~\cite{hulin_strongly_1990} that captures this Griffiths physics and reproduces all essential features of our numerical data. The model [Fig. \ref{fig:rcscaling} (a)] consists of a chain of resistors with a distribution $P(R) \sim R^{-\tau}$, each connected to the ground by a capacitor with a constant capacitance $C$. A power-law distribution of resistances can arise naturally in the physical system, as follows.
Suppose the resistance is dominated by randomly distributed but identical tunneling barriers, such that each site has a probability $p$ of being a barrier. The probability of finding a string of $N$ consecutive barriers is then $p^N$. Standard semiclassical arguments suggest that, if the tunneling rate through any barrier of height $W$ is $e^{-W}$, then the tunneling rate through a string of $N$ barriers scales as $e^{-N W}$; consequently, if $R$ is the (dimensionless) resistance of a single barrier, the resistance of a string of $N$ barriers is $R^N$. Together, these observations imply that the distribution of resistances must satisfy the relation $P[R^N] \simeq \{P[R]\}^N$, and hence that $P[R] \sim R^{-\tau}$ for some $\tau$ whose value depends on microscopic details. 

We now relate this RC model to our numerical results. To this end, we note that, for $1 < \tau < 2$, the average resistance of the chain in the dc limit $ \bar{\rho}(\omega = 0) = \int_{0}^{\infty} P(R) R dR$ is divergent. At any nonzero frequency, however, the capacitors have a finite impedance $1/(\omega C)$, and a resistor with $R_i \gg 1/(\omega C)$ disconnects the circuit into separate blocks. Connecting to our arguments from the previous section, the average size of such blocks is given by $d(\omega) \sim \omega^{-\beta}$, where $\beta$ is again the exponent of subdiffusive relaxation, as for the quantum model; probed on a timescale $t$, the circuit consists of disconnected regions of size $d(1/t)$; thus, the return probability scales as $C_{zz}(t) \simeq [d(1/t)]^{-1} \sim (1/t)^\beta$. 

We can relate $\beta$ to $\tau$ using the following argument. For $ 1< \tau < 2$, the resistance of a block of size $d$ is dominated by its largest \emph{expected} resistor $R_d$, which is given by the criterion $ d \int_{R_d}^\infty P(R) dR = 1$ yielding $R_d \simeq d^{1/(\tau - 1)}$. The conductivity of such a block is given by $\sigma_d \simeq d/R_d$, and through the Einstein relation, is related to the time of diffusion $t_d$ across such a block, $t_d \sim d^2 / \sigma_d \sim d^{-\tau/(1-\tau)} $. At any frequency $\omega$, the size of independent blocks $d(\omega)$ is such that the diffusion time satisfies $t_d \simeq 1/\omega$. This yields $d(\omega) \sim \omega^{- (1-1/\tau)}$, i.e., $\beta = 1-1/\tau$. For $\tau > 2$, the distribution $P(R)$ yields a well defined (length independent) average resistance, and the Einstein relation yields $d(\omega) \sim \omega^{-1/2}$, or $\beta = 1/2$. These results concur with previous more rigorous analyses \cite{hulin_strongly_1990,Machtarandomwalk}. 

The conductivity exponent $\alpha$ is determined by reinstating frequency dependence in the result $\sigma_d \simeq d/R_d$; we find $\alpha = 2/\tau - 1$ and $\alpha = 0$ for $ 1 < \tau < 2$ and $\tau \geq 2$ respectively. Combining this with the previous result for $\beta$, we readily find that the scaling relation $\alpha + 2 \beta = 1$ is satisfied for all values of $\tau$. Next, we calculate the scaling of the width of the resistivity distribution. We again note that segments of the penetration depth $d(\omega)$ behave independently of one another and there are $L/d(\omega)$ such segments in a system of length $L$. 
The width of the resistivity distribution can then be shown (see Supplemental Information) to satisfy $\Delta\rho(\omega) \approx \Delta R(\omega) /\sqrt{L}$, where $\Delta R(\omega)$ is the width of the distribution of single resistors; $[\Delta R(\omega)]^2 \equiv \int_0^{R_d} P(R) (R - \bar{R})^2 dR \sim \omega^{1-3/\tau}$. This yields $\alpha' = 3/2\tau-1/2$ which is slightly greater than $\alpha$ in the range of the Griffiths phase $1 < \tau < 2$ (this is reflected in the numerics for the XXZ model for $h_\text{max} \alt 2.5$, see inset in Fig.~\ref{fig:scaling} (c)), and slightly beyond into the diffusive phase up to $\tau = 3$.  

We have verified these scaling arguments by numerically solving the RC model [Fig.~\ref{fig:rcscaling}] in both the subdiffusive and diffusive regime and find good agreement with the analytical predictions. In particular, we calculate the mean and width of the finite frequency resistivity distribution and find that their asymptotic form is a power law, see (b). To test the relation $\alpha=2/\tau-1$, we plot $(\alpha+1)\tau/2$ as a function of $\tau$, (c), and confirm it to be constant in the subdiffusive regime $1 < \tau < 2$, while it increases linear in the diffusive regime $\tau > 2$, where $\alpha=0$.

\emph{Conclusions}. In this work we have numerically established the following facts about the delocalized phase near the MBL transition in the disordered XXZ chain. (1) The conductivity vanishes at low frequencies with the power law $\sigma(\omega) \sim \omega^\alpha$. (2) Spin transport is subdiffusive, and the return probability at long times decays as $C^i_{zz}(t) \sim 1/t^\beta$, with the saling relation $\alpha + 2\beta = 1$. As the localized phase is approached, $\beta \rightarrow 0$, and $\alpha \rightarrow 1$ while as the diffusive phase is approached $\beta \rightarrow 1/2$ and $\alpha \rightarrow 0$. (3) The distribution of resistivities of a fixed-sized sample grows increasingly broad at low-frequencies, and the width of this distribution diverges as a power law with exponent $\alpha' > \alpha$ at low frequencies. The distribution of resistivities becomes scale-free and presumably power-law in the d.c. limit. These general observations allow us to identify the phase as a Griffiths phase. We 
also derived the central scaling relation $\alpha + 2 \beta = 1$ postulating a scaling form of the spin-diffusion propagator. We showed that a phenomenological, classical RC model allows us to capture the various features of the Griffiths phase in a simple manner. Our predictions can be directly tested in experiments with ultracold atoms in disordered potentials~\cite{billy_direct_2008,roati_anderson_2008,demarco1,demarco2,errico_observation_2014,bakr_quantum_2009,sherson_single-atom-resolved_2010,tilman_2014}, polar molecules~\cite{yan_observation_2013}, nitrogen-vacancy centers in diamond~\cite{doherty_nitrogen-vacancy_2013}, and thin films~\cite{Ovadyahu2012,ovadyahu_2015}.
Two intriguing aspects of the Griffiths phase that remain to be addressed in future work are: (i) whether it is ergodic; and (ii) whether any such phase exists in more than one dimension, where single local bottlenecks cannot block global transport.

\paragraph*{Acknowledgments} The authors acknowledge discussions with 
 D. Abanin, E. Altman, A. Amir, I. Bloch, I. Gornyi, D. Huse, L. Ioffe, I. Lerner, M. Lukin, I. Martin, A. Mirlin, R. Modak, A. Pal, A. Polkovnikov, and N. Yao. 
The authors acknowledge support from Harvard Quantum Optics Center, Harvard-MIT CUA, ARO-MURI Quism program, ARO-MURI on Atomtronics, as well as the Austrian Science Fund (FWF) Project No. J 3361-N20.

\end{document}


\newcommand{\breite}{1.0} 

\newtheorem{prop}{Proposition}
\newtheorem{cor}{Corollary}

\newcommand{\be}{\begin{equation}}
\newcommand{\ee}{\end{equation}}

\newcommand{\bea}{\begin{eqnarray}}
\newcommand{\eea}{\end{eqnarray}}

\newcommand{\Reals}{\mathbb{R}}     
\newcommand{\Com}{\mathbb{C}}       
\newcommand{\Nat}{\mathbb{N}}       

\newcommand{\id}{\mathbbm{1}}    

\newcommand{\Real}{\mathop{\mathrm{Re}}}
\newcommand{\Imag}{\mathop{\mathrm{Im}}}

\def\O{\mbox{$\mathcal{O}$}}   
\def\F{\mathcal{F}}			
\def\sgn{\text{sgn}}

\newcommand{\deo}{\ensuremath{\Delta_0}}
\newcommand{\dea}{\ensuremath{\Delta}}
\newcommand{\ak}{\ensuremath{a_k}}
\newcommand{\ad}{\ensuremath{a^{\dagger}_{-k}}}
\newcommand{\sx}{\ensuremath{\sigma_x}}
\newcommand{\sz}{\ensuremath{\sigma_z}}
\newcommand{\spl}{\ensuremath{\sigma_{+}}}
\newcommand{\smi}{\ensuremath{\sigma_{-}}}
\newcommand{\alk}{\ensuremath{\alpha_{k}}}
\newcommand{\bk}{\ensuremath{\beta_{k}}}
\newcommand{\ok}{\ensuremath{\omega_{k}}}
\newcommand{\vd}{\ensuremath{V^{\dagger}_1}}
\newcommand{\vi}{\ensuremath{V_1}}
\newcommand{\vo}{\ensuremath{V_o}}
\newcommand{\zc}{\ensuremath{\frac{E_z}{E}}}
\newcommand{\xc}{\ensuremath{\frac{\Delta}{E}}}
\newcommand{\xd}{\ensuremath{X^{\dagger}}}
\newcommand{\aok}{\ensuremath{\frac{\alk}{\ok}}}
\newcommand{\tpw}{\ensuremath{e^{i \ok s }}}
\newcommand{\tpe}{\ensuremath{e^{2iE s }}}
\newcommand{\tmw}{\ensuremath{e^{-i \ok s }}}
\newcommand{\tme}{\ensuremath{e^{-2iE s }}}
\newcommand{\epls}{\ensuremath{e^{F(s)}}}
\newcommand{\emis}{\ensuremath{e^{-F(s)}}}
\newcommand{\epl}{\ensuremath{e^{F(0)}}}
\newcommand{\emi}{\ensuremath{e^{F(0)}}}

\newcommand{\mkcomm}[1]{{\color{red}MK: #1}}

\newcommand{\lr}[1]{\left( #1 \right)}
\newcommand{\lrs}[1]{\left( #1 \right)^2}
\newcommand{\lrb}[1]{\left< #1\right>}
\newcommand{\nbt}{\ensuremath{\lr{ \lr{n_k + 1} \tmw + n_k \tpw  }}}

\newcommand{\om}{\ensuremath{\omega}}
\newcommand{\dw}{\ensuremath{\Delta_0}}
\newcommand{\wbp}{\ensuremath{\omega_0}}
\newcommand{\dv}{\ensuremath{\Delta_0}}
\newcommand{\vbp}{\ensuremath{\nu_0}}
\newcommand{\vplus}{\ensuremath{\nu_{+}}}
\newcommand{\vminus}{\ensuremath{\nu_{-}}}
\newcommand{\wplus}{\ensuremath{\omega_{+}}}
\newcommand{\wminus}{\ensuremath{\omega_{-}}}
\newcommand{\uv}[1]{\ensuremath{\mathbf{\hat{#1}}}} 
\newcommand{\abs}[1]{\left| #1 \right|} 
\newcommand{\avg}[1]{\left< #1 \right>} 
\let\underdot=\d 
\renewcommand{\d}[2]{\frac{d #1}{d #2}} 
\newcommand{\dd}[2]{\frac{d^2 #1}{d #2^2}} 
\newcommand{\pd}[2]{\frac{\partial #1}{\partial #2}} 
\newcommand{\pdd}[2]{\frac{\partial^2 #1}{\partial #2^2}} 
\newcommand{\pdc}[3]{\left( \frac{\partial #1}{\partial #2}
 \right)_{#3}} 
\newcommand{\ket}[1]{\left| #1 \right>} 
\newcommand{\bra}[1]{\left< #1 \right|} 
\newcommand{\braket}[2]{\left< #1 \vphantom{#2} \right|
 \left. #2 \vphantom{#1} \right>} 
\newcommand{\matrixel}[3]{\left< #1 \vphantom{#2#3} \right|
 #2 \left| #3 \vphantom{#1#2} \right>} 
\newcommand{\grad}[1]{\gv{\nabla} #1} 
\let\divsymb=\div 
\renewcommand{\div}[1]{\gv{\nabla} \cdot #1} 
\newcommand{\curl}[1]{\gv{\nabla} \times #1} 
\let\baraccent=\= 

\renewcommand{\thesection}{S\arabic{section}}
\renewcommand{\theequation}{S\arabic{equation}}
\renewcommand{\thetable}{S\arabic{table}}
\renewcommand{\thefigure}{S\arabic{figure}}
\renewcommand{\thepage}{S\arabic{page}}

\title{Anomalous diffusion and Griffiths effects near the many-body localization transition - Supplemental Information}

\author{Kartiek Agarwal}
\affiliation{Physics Department, Harvard University, Cambridge, Massachusetts 02138, USA}
\author{Sarang Gopalakrishnan}
\affiliation{Physics Department, Harvard University, Cambridge, Massachusetts 02138, USA}
\author{Michael Knap}
\affiliation{Physics Department, Harvard University, Cambridge, Massachusetts 02138, USA}
\affiliation{ITAMP, Harvard-Smithsonian Center for Astrophysics, Cambridge, MA 02138, USA}
\author{Markus M\"uller}
\affiliation{The Abdus Salam International Center for Theoretical Physics, Strada Costiera 11, 34151 Trieste, Italy}
\author{Eugene Demler}
\affiliation{Physics Department, Harvard University, Cambridge, Massachusetts 02138, USA}

\date{\today}
\maketitle

\section{XXZ Model}

\subsection{Scaling form of $G(k,\omega)$}

The propagator $G(x,t)$ describes the decay of a spin-density configuration $\delta s (x,t_0)$ imprinted at an initial time $t_0$, that is, $\delta s(x , t) = \int dx' \int_{t_0}^t dt' G(x-x',t-t') \delta s(x,t')$. For instance, in the case of regular diffusion, its Laplace Transform $G(k,\omega) = 1/[-i \omega + D k^2]$ is a simple pole that describes exponential relaxation of a spin-density mode with wave-vector $k$ on a time-scale $1/D k^2$. 

In this section we derive the scaling relation $\alpha + 2\beta = 1$ by postulating that the spin density propagator exhibits a scaling form $G(k, \omega) \sim \omega^{-1} \tilde{\phi} (k / \omega^\beta)$, that is, distance and time are related exclusively through a single exponent $\beta$, or equivalently, a dynamical exponent $z = 1/\beta$. The presence of a smoothly varying dynamical exponent is reminiscent of the zero-temperature Griffiths phase predicted in random magnets (see Ref.~\cite{motrunich2001dynamics} and the references therein). We expect that such a correspondence should emerge in the limit $\omega \ll J = 1$, $k \ll 2\pi$. Now, the total spin conservation requires that the function $\tilde{\phi} (k/\omega^\beta)$ approaches a finite value in the $k \rightarrow 0$ limit, such that $G(k \rightarrow 0,\omega) \sim 1/\omega$ and its integral over time is constant. An inverse Laplace-transform yields $G(x,t) \sim t^{-\beta} \phi(x/t^\beta)$, which, in the limit $x \rightarrow 0$ exhibits the time 
dependence of the observed local return probability $C_{zz} (t) \sim 1/t^\beta$. 

While the propagator $G(k,\omega)$ describes the decay of a spin density wave with momentum $k$ given an initial description of the spin-density, the dynamical polarizability $\chi(k,\omega)$ describes the response of the spin-density to a \emph{magnetic field} that is switched on \emph{slowly} from $t = -\infty$ to $t = 0$ (as opposed to directly imprinting a density ripple). The two quantities can be shown to be related through the static polarizability $\chi_0 (k)$ via the relation $\chi(k,\omega) = \chi_0 (k) [ 1 + i\omega G(k,\omega)]$. And finally, the dynamical structure factor is given by $S(k,\omega) = \coth{(\omega/2T)} \, \chi''(k,\omega)$, or, in the infinite-temperature limit, $S(k,\omega) = (2T/\omega) \, \chi''(k,\omega)$. Using these relations, we arrive at the form $S(k,\omega) \sim \omega^{-1} h(k/\omega^\beta)$, provided we assume $T \chi_0 (k)$ has a finite $k \rightarrow 0$ limit: this is justified because a high-temperature system with short-range interactions 
is unexpected to become incompressible. Using the conservation of magnetization, we have $\sigma (k \rightarrow 0, \omega) \sim \omega^2 \partial_k^2 S(k \rightarrow 0,\omega)$. The result $\sigma( k \rightarrow 0, \omega) \sim \omega^{1 - 2 \beta}$ follows directly, and yields the scaling relation $\alpha + 2\beta = 1$.  

\subsection{Finite Size scaling}

\begin{figure*}[t]
\includegraphics[width = 7in]{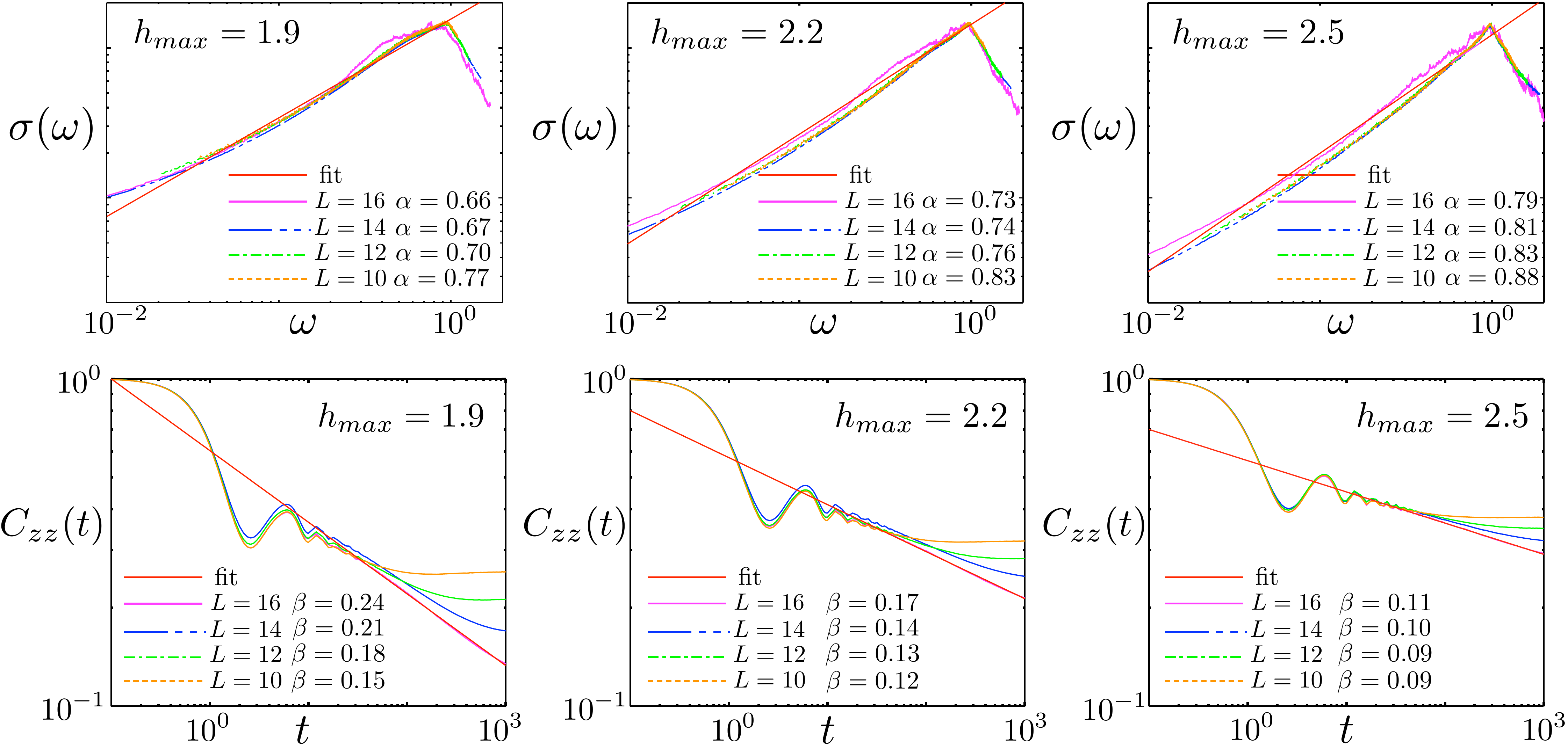}
\caption{Conductivity $\sigma(\omega)$ and return probability $C_{zz} (t)$ shown for $h_{max} = 1.9,2.2,2.5$, $J_z = 8$, and system sizes $L = 10,12,14,16$. The fit interval is $[t_{min},t_{max}]$ for $C_{zz}(t)$ and $[2\pi/t_{max},2\pi/t_{min}]$ for $\sigma(\omega)$, with $t_{min} = 20$ for all $L$ and $t_{max} = 100,400,800$ for $L = 12,14,16$ respectively (fit function is only shown for $L=16$). The power-law range for $L = 10$ is limited and was not used for extrapolation of exponents.}
\label{fig:finsize}
\end{figure*}

\begin{figure}[t]
\includegraphics[width = 2.5in]{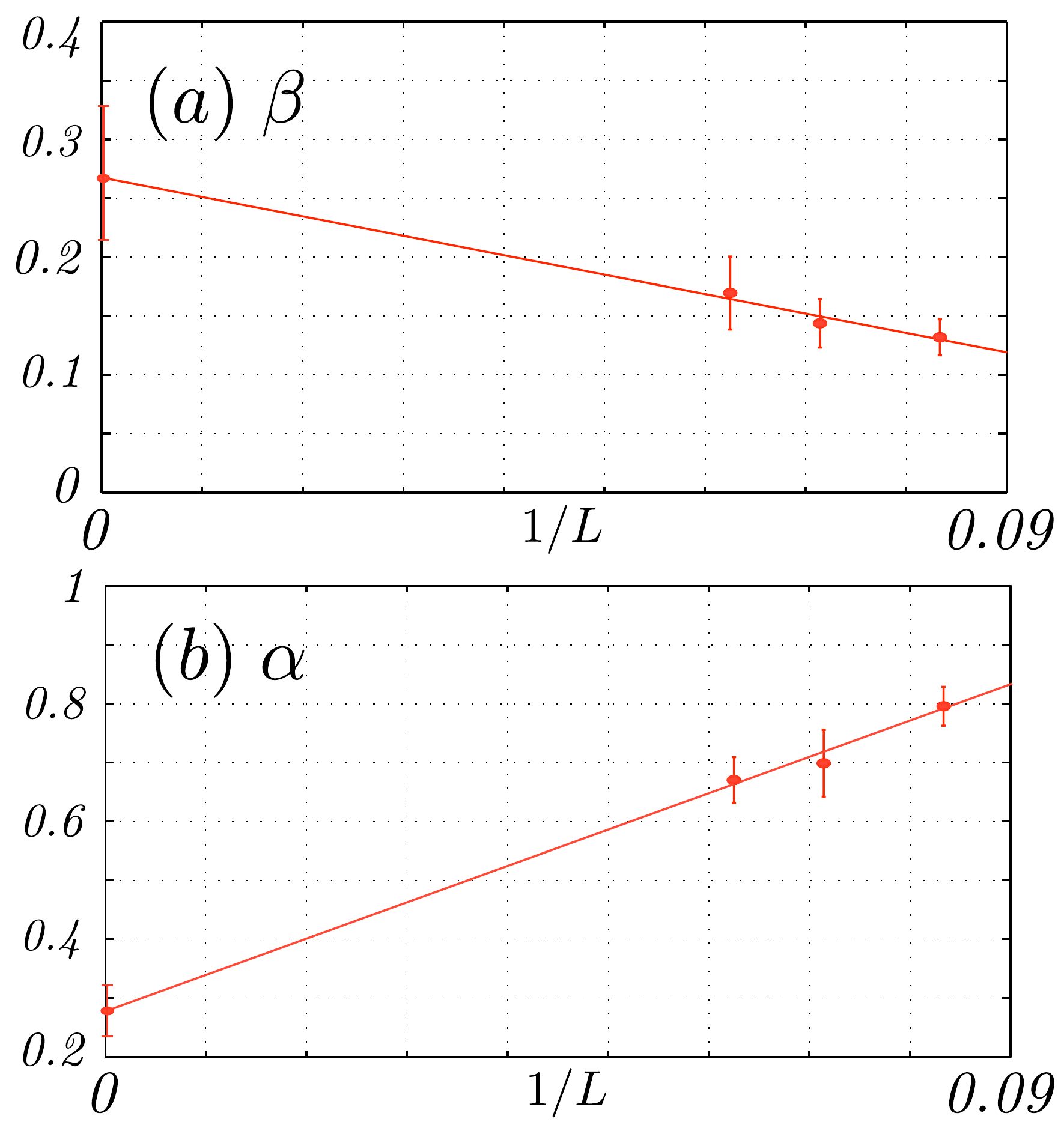}
\caption{Extracted values of $\alpha$ and $\beta$ are plotted against the inverse system size $1/L$ for $h_{\text{max}} = 2.2$ and $J_z = 1$. The solid line shows the extrapolation of these exponents to $L = \infty$.}    
\label{fig:extrap}
\end{figure}

The exponents $\alpha$ and $\beta$ corresponding to the conductivity $\sigma(\omega) \sim \omega^\alpha$ and return-probability $C_{zz} (t) \sim 1/t^\beta$ were extracted from systems of size $L = 12,14,16$ (see Fig.~\ref{fig:finsize} and \ref{fig:extrap})for various $h_{max}$ and extrapolated to the $L \rightarrow \infty$ limit using the extrapolation $\alpha(L) \sim \alpha_{\infty} + x/L$ and $\beta(L) \sim \beta_{\infty} +  y/L$. Here, $x$ and $y$ are undetermined coefficients, and $\alpha_{\infty}$ and $\beta_{\infty}$ correspond to the $L \rightarrow \infty$ values of the exponents, which have been plotted in Fig.~1 of the main text, with error bars indicating $1\sigma$ confidence intervals.  

\begin{figure}[t]
\includegraphics[width = 2.5in]{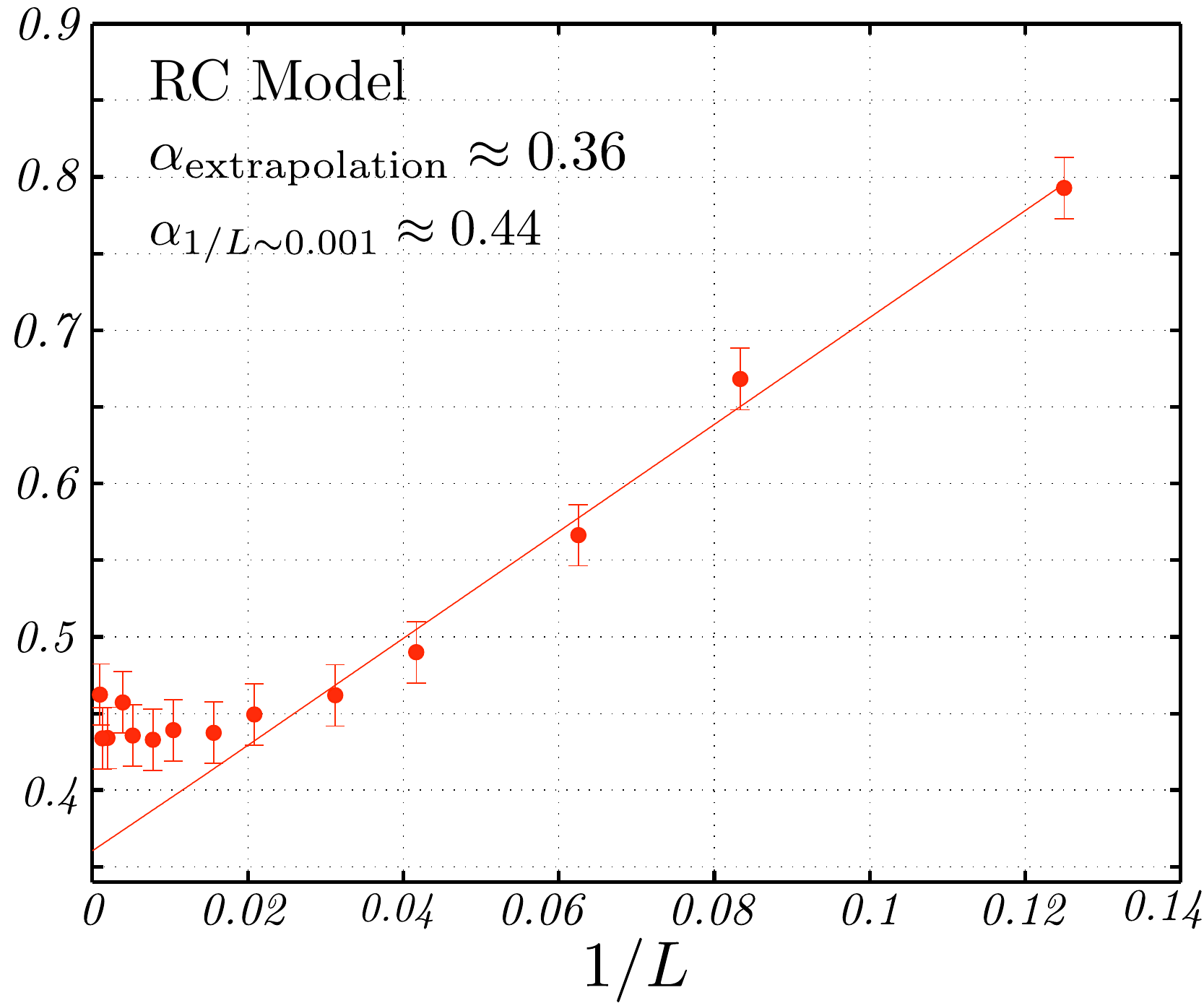}
\caption{Value of the conductivity exponent $\alpha$ plotted against the inverse system size $1/L$ for the RC model, for $\beta = 0.25$. The solid line shows the extrapolation of these exponents to $L = \infty$.}    
\label{fig:rcexp}
\end{figure}

The extrapolation scheme is based on the observed finite-size scaling of the exponents in the RC model (see Fig.~\ref{fig:rcexp}) -- for small system sizes, the exponent scales with system size as $1/L$ but eventually stops scaling for large enough system sizes. The extrapolation to $L \rightarrow \infty$ based on $1/L$ scaling over small system sizes, however, is seen to produce a reasonable estimation of the $L \rightarrow \infty$ value of the exponent. 

\begin{figure}[t]
\includegraphics[width = 2.5in]{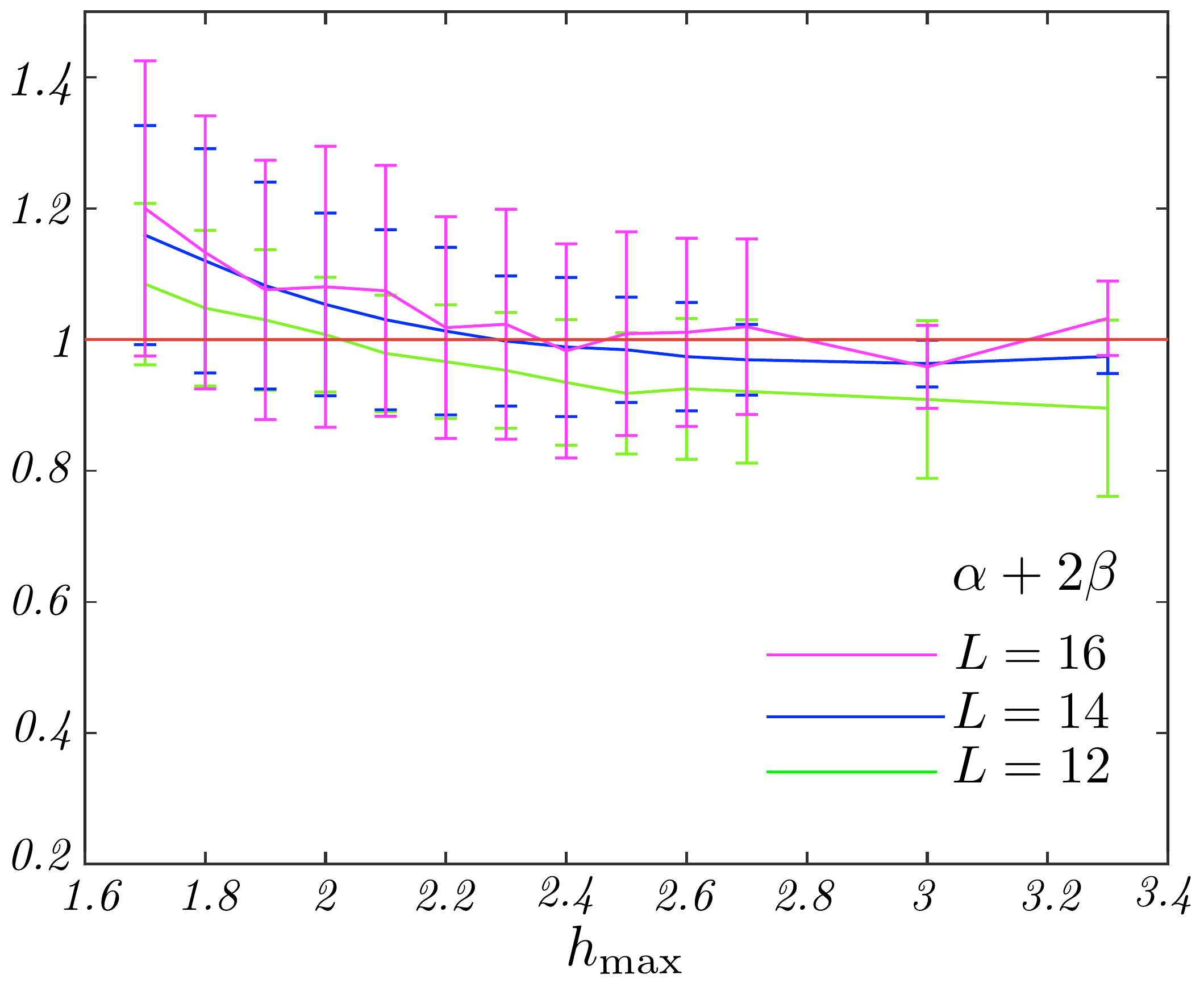}
\caption{$\alpha + 2\beta$ plotted against $h_\text{max}$ for $J_z = 1$, and system sizes $L = 12,14,16$. }    
\label{fig:alphabetasumvsL}
\end{figure}

While the scaling relation $\alpha + 2 \beta =1$ can be seen to hold approximately for the extrapolated exponents in Fig. 1 of the main text, it is also seen to hold for fixed system size $L = 12,14,16$ in Fig.~\ref{fig:alphabetasumvsL}. 

\begin{figure}[t]
\includegraphics[width = 3in]{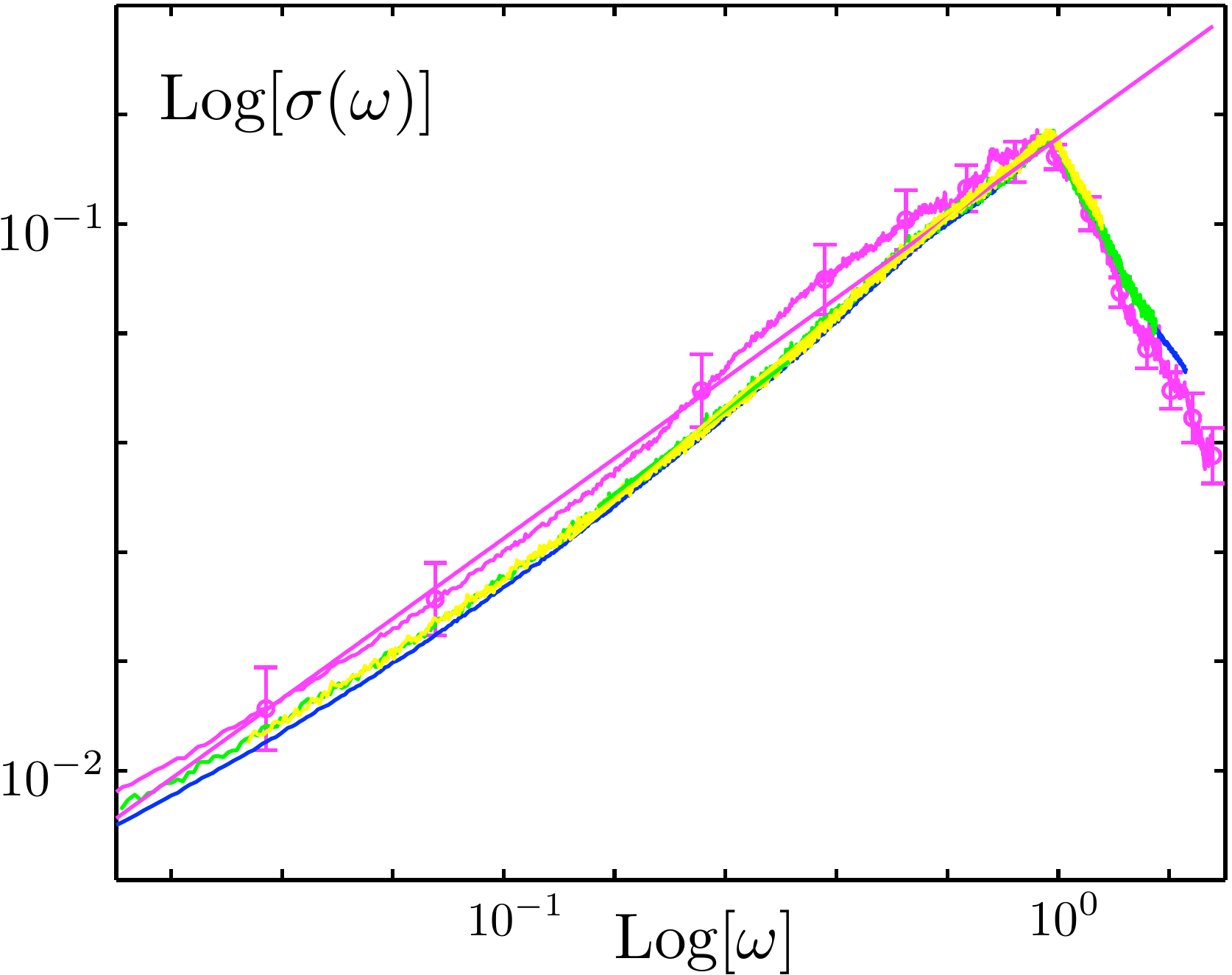}
\caption{Optical conductivity is plotted against frequency for $h_\text{max} = 2.2$, $J_z = 0.8$, and system sizes $L = 10,12,14,16$. The error bars on $L = 10,12,14$ are negligible and not plotted. The error bars on $L = 16$ are large because of the smaller number of samples simulated, and of the order of the deviation observed from the result of smaller system sizes.}    
\label{fig:L16error}
\end{figure}

It must also be noted that the $L = 16$ simulations were carried out over a much smaller set of $\sim 20$ samples, and the deviations in the conductivity measurement from $L = 10,12,14$ system sizes falls within the scope of the larger error bars (see Fig.~\ref{fig:L16error}). 

\subsection{Numerical Evaluation of Conductivity}

\begin{figure}[h]
\includegraphics[width = 3in]{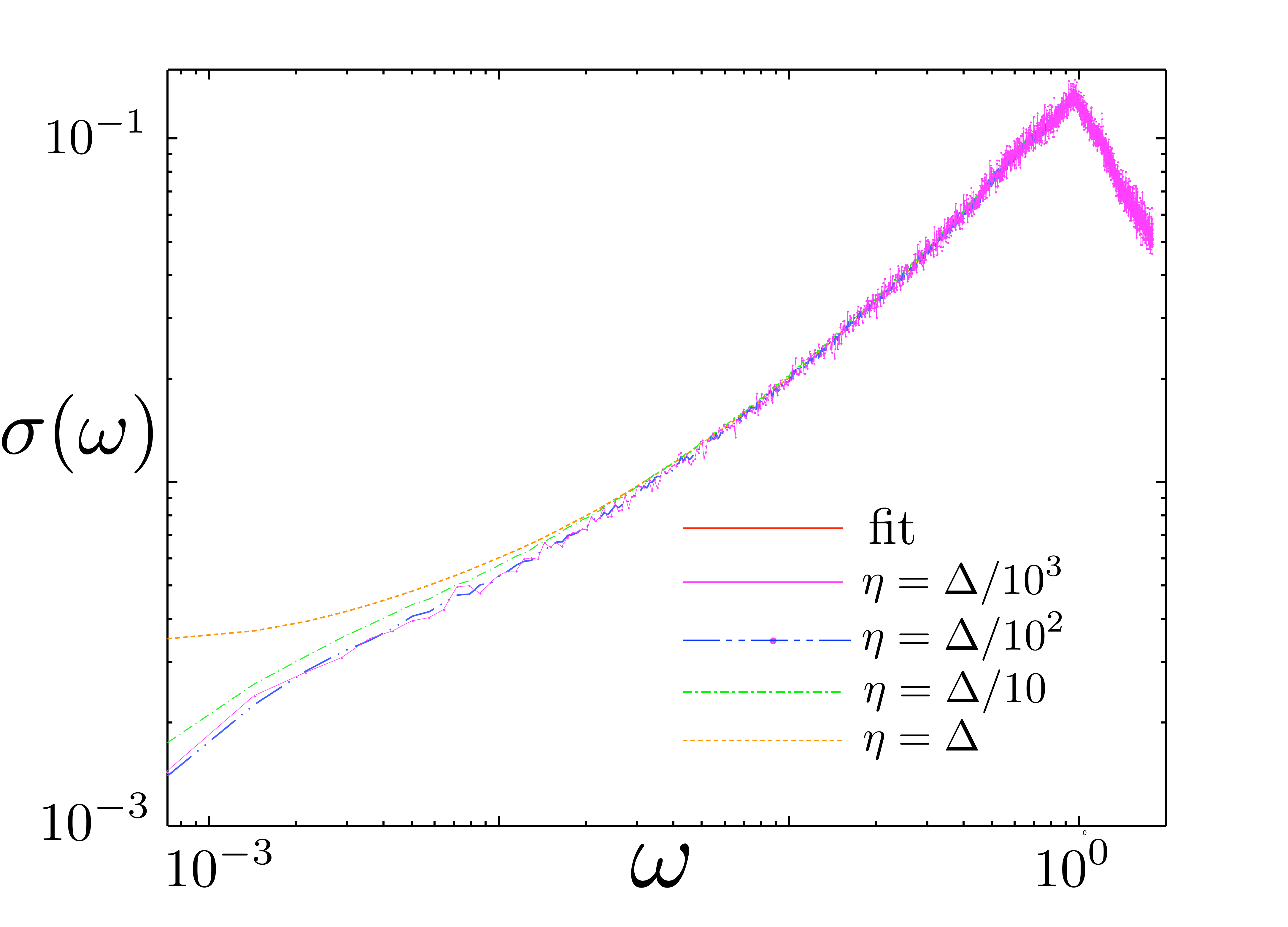}
\caption{Conductivity for different values of the width $\eta$ of the Lorentzian representing the $\delta$-function in the Kubo formula. $\Delta$ is the average level spacing. A convergence can be seen in the limit of small $\eta$.}
\label{fig:lorentzwidth}
\end{figure}

The numerical evaluation of the conductivity using the Kubo-formula in the main text requires making a choice on the precise form of the $\delta$-function -- for our purposes, we typically use a Lorentzian of width $\eta \sim \Delta/10^2$, centered at $\omega_{mn}$, where $\Delta$ is the average level spacing of the system, determined approximately by $\Delta \sim h_{max} \sqrt{L} / \binom{L}{L/2}$. In Fig. \ref{fig:lorentzwidth}, the conductivity is plotted for a few different values of $\eta$. A convergence is seen as $\eta$ approaches smaller values, indicating the validity of our choice of $\eta \sim \Delta/10^2$. 

\subsection{Typical Decay vs. Average Decay}

\begin{figure}[t]
\includegraphics[width = 3in]{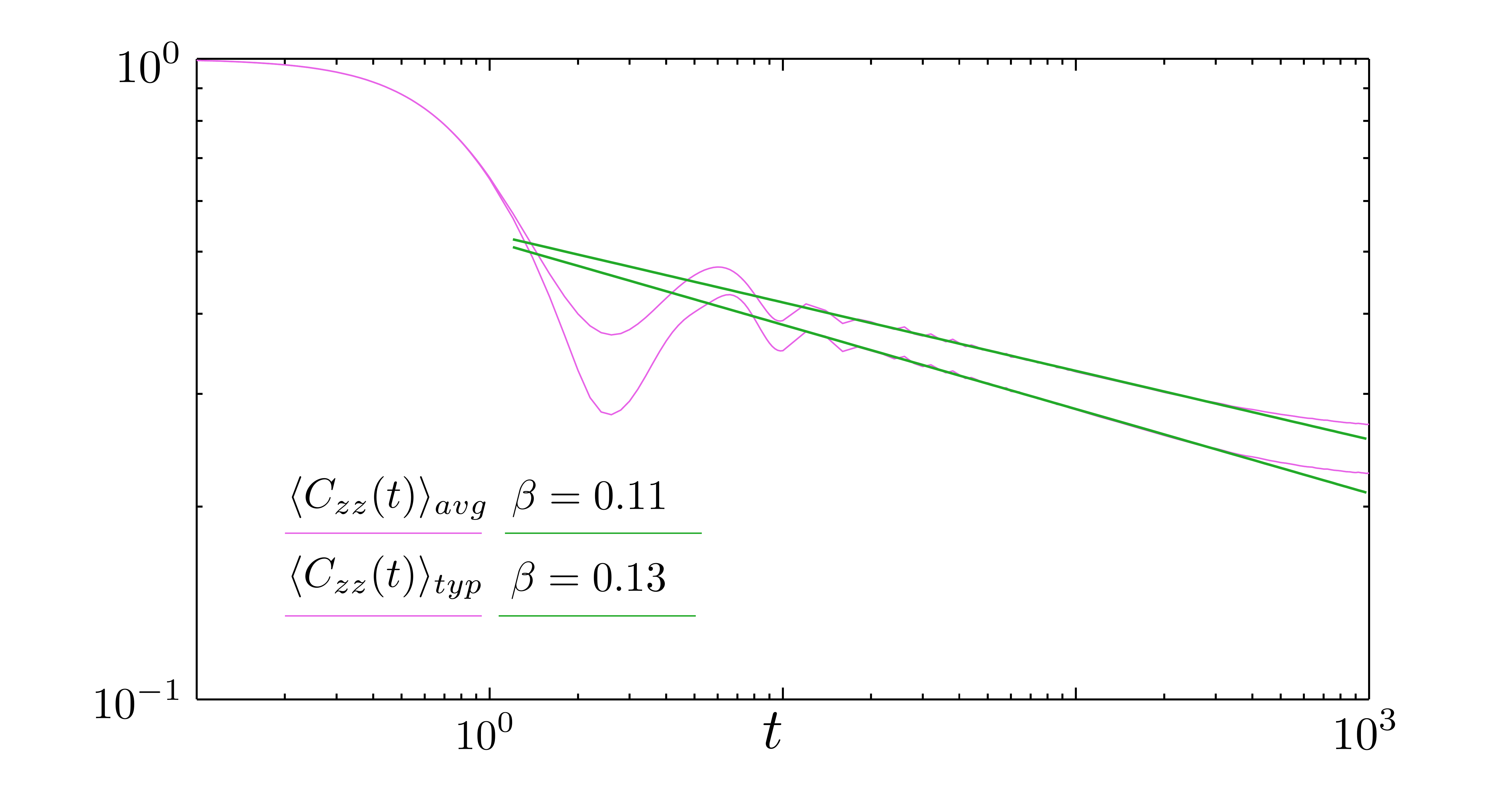}
\caption{Both typical and average return probability show similar power-law behavior. Plots are for $J_z = 0.8$, $h_\text{max} = 2.3$. }
\label{fig:typical}
\end{figure}

\begin{figure}[t]
\includegraphics[width = 3in]{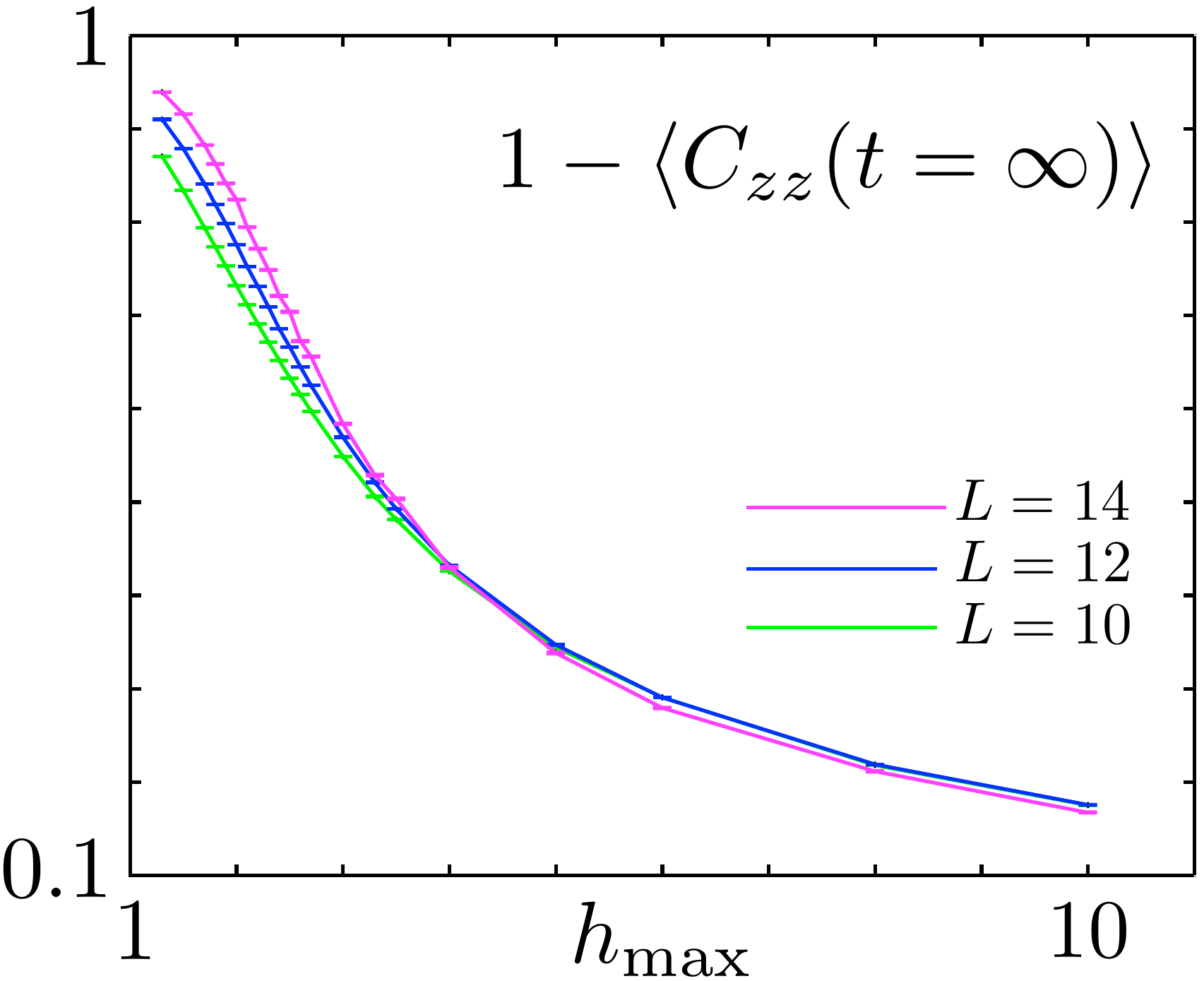}
\caption{ The infinite-time return probability $C_{zz} (t = \infty)$, as a function of disorder $h_{max}$ for $J_z = 1$.}
\label{fig:finale}
\end{figure}

It is important to note that both average (linearly averaged) and typical (logarithmically averaged) quantities show a power law behavior. This crucial information allows us to distinguish between two alternative pictures of relaxation in our system -- i) the local density \emph{typically} relaxes diffusively, but on rare sites, it relaxes sub-diffusively owing to a proximity to an insulating region. This gives rise to a sub-diffusive behavior of relaxation, on \emph{average}, but the \emph{typical} relaxation will still be diffusive vs. ii) the local density at any site relaxes sub-diffusively, because the relaxation process is determined by possibly faraway insulating regions. Thus, both \emph{typical} and \emph{average} relaxation is sub-diffusive. In our system, we find the second case to be true, and this is crucial to explaining the scaling relations and power-law dynamics that we observe. Fig. \ref{fig:typical} shows the average return probability, $\sum_{\{D\}} C^i_{zz} (t) / N_D$, 
where $\{D\}$ is the set of quenched disorder configurations we perform averaging over, and $N_D$ is the number of disorder realizations) and typical return probability, $\exp [ \sum_{\{D\}} \log C^i_{zz} (t) / N_D ]$. 

\subsection{Infinite-time Decay}

In Fig. (\ref{fig:finale}) we plot the infinite-time return probability $C_{zz} (t = \infty)$, as a function of disorder $h_{max}$ for the Heisenberg case, $J_z = 1$ across the MBL transition. Fig.~(\ref{fig:finale}) shows that, as $\beta \rightarrow 0$, the infinite-time return probability becomes appreciably greater than the inverse system size, marking the onset of the MBL phase. 

\section{RC Model}

\subsection{Width of resistivity distribution in the RC Model}

It is argued in the main text that the description of a finite system of length $L$ as composed of independent resistive blocks of length $d(\omega)$ gives rise to a width of the resistivity distribution that satisfies $\Delta \rho(\omega) \approx \Delta R / \sqrt{L} $. Here we substantiate this claim. 

The effective distribution of resistors in a block of size $d(\omega)$ is given by $P(R) = 1/R^\tau$ with a cut-off at $R_d(\omega)$, for which we can calculate a mean $\bar{R}(\omega)$ and a width $\Delta R(\omega)$. The finiteness of the penetration depth provides a cut-off to this effective distribution, which makes it well-behaved. This allows us to calculate the resistivity (which is the arithmetic mean of such $d(\omega)$ resistors) distribution of the block using the central-limit theorem : the mean $\bar{\rho}_d(\omega) \approx \bar{R} (\omega)$ and width $\Delta \rho_d (\omega) \approx \Delta R (\omega) / \sqrt{d(\omega)}$. Alternatively, one can calculate the mean and width of the distribution of the conductivity $\sigma_d(\omega)$ for such a block, using $\bar{\sigma}_d (\omega) \approx 1/\bar{\rho}_d (\omega)$ and $\Delta \sigma_d (\omega) \approx \Delta \rho_d (\omega) / [\bar{\rho}_d (\omega)]^2$. 

These blocks of the size of the penetration depth behave independently, and consequently, the total conductivity of the system of size $L$ is an arithmetic mean of these blocks. The mean $\bar{\sigma}_L (\omega)$ and width $\Delta \sigma_L (\omega) $ are recovered using another application of the central-limit theorem noting that the number of such blocks is $L / d(\omega)$. Following through with this calculation, one finds, $\Delta \sigma_L (\omega) \approx \Delta R (\omega) / [ \bar{R} (\omega) ]^2 \sqrt{L} $ and $\bar{\sigma}_L (\omega) \approx 1/\bar{R}(\omega)$ (the latter result is derived in the main text as well). 
To finally obtain the distribution of the resistivity $\rho_L (\omega)$ of the system of length $L$, we use $\Delta \rho_L (\omega) \approx \Delta \sigma_L (\omega) / [\bar{\sigma}_L (\omega)]^2$, and $\bar{\rho}_L (\omega) \approx 1/\bar{\sigma}_L (\omega)$. This finally yields the result mentioned in the main text, that the frequency dependence of the width of the resistivity $\rho(\omega)$, is given by $\Delta \rho_L (\omega) \approx \Delta R(\omega) / \sqrt{L} $.

\subsection{RC model simulations}

As mentioned in the main text, the RC circuit is to be imagined as a series of non-resistive, metallic grains that are connected to one-another though resistors that are analogues of Miller-Abrahams resistors~\cite{miller_abrahams} $R_i$. The current $I_i$ between the grain $i,i+1$ is given by $ (\mu_i - \mu_{i+1})/R_i$, where $\mu_i$ is the chemical potential on the grain $i$~\cite{ambegaokar_hopping,amir2014universal}. These metallic grains are short-circuited through a capacitance $C$, and the charge $q_i$ on the capacitors is given by the total electro-chemical potential on each grain, $(V_i + \mu_i)/C$ where $V_i = V_o e^{i\omega t + i q x_i}$ is the external potential applied to measure the conductivity of the circuit. The currents $I_i$ and $I_{i-1}$ flowing into these grains thus satisfy the relation $I_{i-1} - I_i = \dot{q}_i$. This set of linear equations are solved to obtain the real part of the conductivity (in the $q\rightarrow 0$ limit) which is the time-averaged power dissipated in the 
circuit, given by $\sigma(q\rightarrow 0, \omega) = \sum_i I_i^2 (q) R_i/ L (V_o q)^2$.

%